\documentclass[showpacs,pra,twocolumn]{revtex4}
\usepackage{amssymb}
\usepackage{graphicx}
\usepackage{amsmath}
\usepackage[sort&compress]{natbib}
\usepackage{color}
\usepackage{bm}
\usepackage{epsfig}
\usepackage[colorlinks]{hyperref}

\begin{document}

\date{\today}
\title{Vortex solitons in two-dimensional spin-orbit coupled Bose-Einstein
condensates: effects of the Rashba-Dresselhaus coupling and the Zeeman
splitting}
\author{Hidetsugu Sakaguchi}
\affiliation{Department of Applied Science for Electronics and Materials,
Interdisciplinary Graduate School of Engineering Sciences, Kyushu
University, Kasuga, Fukuoka 816-8580, Japan}
\author{E. Ya. Sherman}
\affiliation{Department of Physical Chemistry, University of the Basque Country UPV-EHU,
48940, Bilbao, Spain}
\affiliation{IKERBASQUE, Basque Foundation for Science, Bilbao, Spain}
\author{Boris A. Malomed}
\affiliation{Department of Physical Electronics, School of Electrical Engineering,
Faculty of Engineering, Tel Aviv University, Tel Aviv 69978, Israel}

\begin{abstract}
We present an analysis of two-dimensional (2D) matter-wave solitons, governed by
the pseudo-spinor system of Gross-Pitaevskii equations with self- and
cross-attraction, which includes the spin-orbit coupling (SOC) in the
general Rashba-Dresselhaus form, and, separately, the Rashba coupling and
the Zeeman splitting. Families of semi-vortex (SV) and mixed-mode (MM)
solitons are constructed, which exist and are stable in free space, as the
SOC terms prevent the onset of the critical collapse and create the
otherwise missing ground states in the form of the solitons. The Dresselhaus
SOC produces a destructive effect on the vortex solitons, while the Zeeman
term tends to convert the MM states into the SV ones, which eventually
suffer delocalization. Existence domains and stability boundaries are
identified for the soliton families. For physically relevant parameters of
the SOC system, the number of atoms in the 2D solitons is limited by $\sim
1.5\times 10^{4}$. The results are obtained by means of combined analytical
and numerical methods.
\end{abstract}

\maketitle

\section{Introduction.}

Recent developments in producing macroscopic ensembles of cold atoms have
greatly extended an experimentally accessible variety of quantum phenomena
in systems of interacting particles, with both repulsion and attraction
between them. Bosonic gases can be optically cooled down to the temperature
of the Bose-Einstein condensation (BEC) \cite{Ketterle02,BEC-books}.
Properties of these condensates, including the strength and sign of
inter-atomic interactions, can be controlled by means of the Feshbach
resonance \cite{Cornish}, which provides a powerful tool for the creation
and control of various phases in the ultracold gases.

Recently, a great deal of interest has been drawn to the experimentally
demonstrated \cite{Lin} ability of binary BEC to emulate the spin-orbit
coupling (SOC) and Zeeman effect, which play a fundamentally important role
in the solid-state physics. Both types of the SOC known in terms of the
solids, which are represented by the Dresselhaus \cite{Dresselhaus} and
Rashba \cite{Rashba} Hamiltonians, can be simulated in BEC, along with the
Zeeman splitting \cite{Campbell}. Basic results produced by recent research
in this field have been summarized in reviews \cite{Zhai2012,reviews}.
Although the true spin of bosonic atoms, such as $^{87}$Rb, used in these
experiments, is zero, the (pseudo)spinor wave function of the binary
condensate has two components, which enables the use of the corresponding
\textit{pseudospin} $1/2$ for emulating the quantum dynamics of fermionic
carriers in solids by means of the bosonic gases. A majority of experimental
works on the SOC \cite{reviews} were dealing with effectively one-dimensional (1D)
settings. Recently, the realization of the SOC in the two-dimensional (2D) geometry was
reported \cite{2D-experiment,Fermi}, which suggests further theoretical
analysis of the 2D settings.

While the (pseudo) SOC in BEC is represented by the linear interaction
between two spatially inhomogeneous components of the condensate (as it must
be, because it emulates the effects from linear quantum mechanics), see Eqs.
(\ref{GPERD1}) and (\ref{GPERD2}) below, the interplay of this linear
coupling with the mean-field nonlinearity of the BEC gives rise to diverse
nonlinear phenomena, such as 1D solitons \cite{soliton},
2D gap solitons \cite{gap-sol}, stripe phases \cite{stripe},
and many others \cite{others}. Further, it is well known that BEC can form
vortex structures \cite{Feynman,Fetter}. Naturally, matter-wave patterns in
the form of single vortices and vortex lattices are nontrivially affected by
the SOC \cite{vortex}. In particular, these works have demonstrated that
specific to the SOC system are 2D composite states in the form of \textit{%
half-vortices} (we call them \textit{semi-vortices} (SVs) below, following
Ref. \cite{Sakaguchi14}), in which one component carries vorticity $S=\pm 1,$
while the other one has $S=0$. The SOC also plays an important role in the
formation of three-dimensional (3D) BEC structures \cite{Ozawa}, including
complex topologically organized modes, such as skyrmions \cite{skyrmion}.

A majority of the above-mentioned works, except for those which were dealing
with 1D bright solitons \cite{soliton}, addressed the binary BEC with the
self-repulsive intrinsic nonlinearity, and cross-repulsion between two
components of the pseudo-spinor wave function. In the case of
self-attraction, a commonly known problem is that 2D and 3D fundamental
(zero-vorticity) solitons, supported by cubic terms, are strongly unstable
in the free space, due to the occurrence of the critical and supercritical
\textit{wave collapse} in the same 2D and 3D settings, respectively \cite%
{collapse}, while vortical solitons are subject to a still stronger
splitting instability \cite{review}. In particular, in the 2D case, the
Gross-Pitaevskii/nonlinear Schr\"{o}dinger equation (GPE/NLSE) with the
cubic self-attraction term gives rise to \emph{degenerate} families of
fundamental \textit{Townes solitons} with $S=0$ \cite{Townes} and their
vortical counterparts with $S\geq 1$ \cite{Minsk}. The degeneracy means that
the entire family has a single value of the norm [see Eq. (\ref{Nmax})
below], which is, as a matter of fact, the value separating collapsing and
decaying solutions, hence the Townes solitons, that play the role of
separatrices between these two types of the dynamical behavior, are
themselves completely unstable. In turn, the degeneracy is a consequence of
the specific scale invariance of the GPE/NLSE in two dimensions.

An unexpected result, which opens novel perspectives for the use of the SOC
in BEC, was recently reported in Ref. \cite{Sakaguchi14} (see also an
extension of the analysis in Refs. \cite{Fukuoka2,Cardoso}): the linear SOC
terms of the Rashba type \cite{Rashba} stabilize 2D free-space solitons, in
the form of the above-mentioned semi-vortices, with $S=0$ and $S=\pm 1$ in
the two components, or in the form of \textit{mixed modes} (MMs), i.e.,
soliton complexes with combinations of terms carrying $S=0$ and $S=+1$ in
one component, and $S=0$, $S=-1$ in the other. The explanation to this
benign effect of the SOC is that this coupling is characterized by an
additional length parameter, that is the spin precession length (typically,
of the order of few micron \cite{Lin,Campbell,Zhai2012,reviews}), inversely
proportional to the SOC strength. Thus, the SOC defines a specific length
scale in the system, thus breaking the above-mentioned scale invariance,
lifting the norm degeneracy of the solitons, and eventually pushing their
norm \emph{below} the threshold necessary for the onset of the collapse.
Being protected against the collapse, the SV and MM solitons enjoy
stabilization and actually introduce a ground state (GS), which is missing
in the scale-invariant 2D GPE/NLSE with the self-attraction \cite%
{Sakaguchi14}. The aptitude of the SOC terms in 2D to suppress the immediate
onset of the critical collapse was also demonstrated in other contexts
(unrelated to solitons) in Refs. \cite{extra} and \cite{Mardonov15}. The
stabilization of 2D solitons in free space is a possibility of obvious
interest to many nonlinear systems beyond the limits of the studies of cold
gases. In particular, it has been demonstrated that 2D spatiotemporal
solitons can be stabilized by means of a similar mechanism in an optics
setting, based on a planar dual-core nonlinear waveguide with temporal
dispersion of the linear coupling between the cores \cite{optics}.

In 3D, the supercritical collapse cannot be suppressed by the SOC terms.
Nevertheless, recent work \cite{HP} has demonstrated that the interplay of
the linear SOC interactions and the cubic self- and cross-attraction gives
rise to two-component 3D solitons of the same two generic types, SVs and
MMs, which are metastable states. Although they cannot be protected against
falling into the supercritical collapse, if subjected to sudden strong
compression, these self-trapped modes are completely stable against small
perturbations.

The stability of the 2D SV and MM solitons was established in Refs. \cite%
{Sakaguchi14,Fukuoka2,Cardoso} for the SOC of the Rashba type, acting in the
combination with the cubic self-attraction, and cross-attraction between the
two components of the binary BEC. Because a physically relevant generic
situation includes a combination of the Rashba and Dresselhaus SOC \cite%
{Campbell}, and the stability of the 2D solitons in free space is quite an
unexpected result, it is relevant to extend the analysis to the full
Rashba-Dresselhaus Hamiltonian, which is one of major objectives of the
present work. This analysis is performed in Section 2, %where we also
%uncover qualitative difference between the SV and MM modes, in terms of the
%anomalous spin-dependent velocity, which can be used to distinguish them
%experimentally
a general conclusion being that the addition of the Dresselhaus interaction
leads to shrinkage of the existence regions for stable 2D solitons,
eventually leading to onset of delocalization. The analysis establishes the
existence boundaries of the SV and MM solitons in the case of the combined
Rashba-Dresselhaus interaction.

Another objective of this work is to consider effects of the Zeeman
splitting on the 2D solitons, which is an obviously relevant problem too, in
terms of both the emulation of the solid-state physics and the BEC dynamics
per se. This problem is addressed by means of combined analytical and
numerical methods, including a variational approximation (VA), in Section 3,
where it is shown that the increase in the Zeeman terms leads to a
transition from the MM solitons to their SV counterparts, and, eventually,
to delocalization. The paper is concluded by Section 4.

\section{Combined effects of the Rashba and Dresselhaus spin-orbit couplings
on 2D solitons}

\subsection{The model and classification of the states}

We begin with the consideration of the action of the synthetic SOC,
including both the Rashba and the Dresselhaus terms, in the 2D space, $%
\left( x,y\right) $. In the mean-field approximation \cite{BEC-books}, the
pseudo-spinor condensate is described by a two-component wave function \cite%
{Zhai2012,reviews}, $\left[ \phi _{+},\phi _{-}\right] ^{\mathrm{T}}$, with
norm
\begin{equation}
N=\iint \left( \left\vert \phi _{+}\right\vert ^{2}+\left\vert \phi
_{-}\right\vert ^{2}\right) dxdy,  \label{N}
\end{equation}%
which is proportional to the total number of atoms in the binary BEC. The
applicability of this approach to SOC settings was demonstrated in many
works \cite{Zhai2012,reviews,mean-field}, including the case of attractive
interactions \cite{mean-field-attraction}. Accordingly, the evolution of the
wave function is governed by the system of coupled GPEs, written here in the
scaled form \cite{Lin,Zhai2012,reviews}:
\begin{gather}
i\frac{\partial \phi _{+}}{\partial t}=-\frac{1}{2}\nabla ^{2}\phi
_{+}-(|\phi _{+}|^{2}+\gamma |\phi _{-}|^{2})\phi _{+}+  \notag \\
\left( \lambda \widehat{D}^{[-]}\phi _{-}-i\lambda _{D}\widehat{D}^{[+]}\phi
_{-}\right) ,  \label{GPERD1} \\
i\frac{\partial \phi _{-}}{\partial t}=-\frac{1}{2}\nabla ^{2}\phi
_{-}-(|\phi _{-}|^{2}+\gamma |\phi _{+}|^{2})\phi _{-}-  \notag \\
\left( \lambda \widehat{D}^{[+]}\phi _{+}+i\lambda _{D}\widehat{D}^{[-]}\phi
_{+}\right) ,  \label{GPERD2}
\end{gather}%
where $\lambda $ and $\lambda _{D}$ are constants of the Rashba and
Dresselhaus SOC ($\lambda \equiv 1$ is fixed below by means of rescaling %
{\cite{Haddad}}), the nonlinear interactions are assumed to
be attractive, $\gamma $ being the relative strength of the cross-attraction
between the two components, while the self-attraction coefficients are
scaled to be $1$, and we introduced operators $\widehat{D}^{[\pm ]}\equiv
\partial /\partial x\pm i\partial /\partial y$. As for parameter $\gamma ,$
it may be adjusted, if necessary by means of the Feshbach resonance. The
applicability of this technique to the BEC under the action of the SOC was
recently analyzed in Ref. \cite{FR}, see also review \cite{Zhai2012}.

The comparison of scaled 2D equations (\ref{GPERD1}) and (\ref{GPERD2}) with
the underlying system of 3D GPEs presented in the physical units readily
shows that the unit length in these equations corresponds to the spatial
scale about $1$ $\mathrm{\mu }$m. Further, by assuming typical values of the
transverse confinement length $\simeq 3$ $\mathrm{\mu }$m and the scattering
length $\sim -0.1$ nm describing the interatomic attraction, we find that $%
N=1$ in the present notation is tantamount to $\simeq 3,000$ atoms.

The spectrum of plane waves generated by the linearized version of Eqs. (\ref%
{GPERD1}), (\ref{GPERD2}), $\phi _{\pm }\sim \exp \left( i\left( \mathbf{k}%
\cdot \mathbf{r}\right) -i\epsilon t\right) $, where $\mathbf{k}$ is the
wave vector, contains two branches,
\begin{equation}
\epsilon _{\pm }=\frac{k^{2}}{2}\pm \sqrt{(1+\lambda _{D}^{2})k^{2}+4\lambda
_{D}k_{x}k_{y}}.  \label{eps}
\end{equation}%
Note that the spectrum is anisotropic in the presence of the Dresselhaus
coupling, $\lambda _{D}\neq 0.$ The anisotropy, even if relatively weak,
qualitatively alters the vortex solitons, and eventually causes their
delocalization, as shown below.

In the absence of the Dresselhaus coupling, 2D self-trapped states
(solitons) of two different types, SVs and MMs, were obtained as stationary
solutions of Eqs. (\ref{GPERD1}),(\ref{GPERD2}) in Refs. \cite{Sakaguchi14}
and \cite{Fukuoka2}. Imaginary-time simulations, as well as an analytical
variational approximation, produced these solutions starting from the
Gaussian \textit{ans\"{a}tze} written in terms of the polar coordinates, $%
\left( r,\theta \right) $:
\begin{eqnarray}
\left( \phi _{+}\right) _{\mathrm{SV}} &=&A_{1}e^{-\alpha_{1}r^{2}-i\mu t},
\notag \\
\left( \phi _{-}\right) _{\mathrm{SV}} &=&A_{2}re^{i\theta }e^{-\alpha
_{2}r^{2}-i\mu t},  \label{SV}
\end{eqnarray}%
\begin{eqnarray}
\left( \phi _{+}\right) _{\mathrm{MM}} &=&B_{1}e^{-\beta _{1}r^{2}-i\mu
t}-B_{2}re^{-i\theta }e^{-\beta _{2}r^{2}-i\mu t},  \notag \\
\left( \phi _{-}\right) _{\mathrm{MM}} &=&B_{1}e^{-\beta _{1}r^{2}-i\mu
t}+B_{2}re^{i\theta }e^{-\beta _{2}r^{2}-i\mu t},  \label{MM}
\end{eqnarray}%
where $\mu $ is the chemical potential, factors $\exp \left( \pm i\theta
\right) $ carry vorticities $S=\pm 1$, while constants $A_{1,2},$ $B_{1,2}$
and $\alpha _{1,2}^{-1/2},$ $\beta_{1,2}^{-1/2}$ represent the amplitudes
and widths of the input.

In the case of $\lambda _{D}=0$, the so obtained SV and MM modes are stable
and produce the system's GS, severally, at $\gamma <1$ and $\gamma >1$.
Precisely at $\gamma =1$, both types are stable and, moreover, they are
embedded into a broader family with an extra free parameter (existing solely
at $\gamma =1$), which makes it possible to link the SV and MM modes \cite%
{Fukuoka2}.

\subsection{Semi-vortices and mixed-mode states}

\begin{figure}[tbp]
\begin{center}
\includegraphics[width=0.45\textwidth]{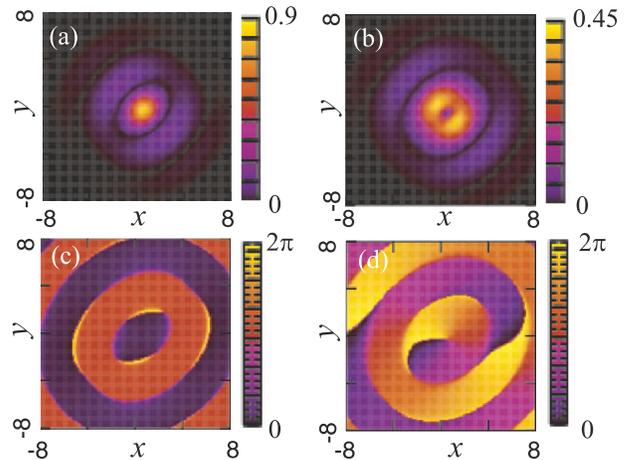}
\end{center}
\caption{Plots of (a) $|\protect\phi _{+}\left( x,y\right),|$ (b) $|\protect%
\phi _{-}\left( x,y\right)|,$ (c) phase of $\protect\phi _{+}(x,y),$ and (d)
phase of $\protect\phi _{-}( x,y) $ for a semi-vortex soliton at $\protect%
\lambda _{D}=0.05,$ $\protect\gamma =0.9,$ and $N=3.5$. It is worth
mentioning that strong deformation of the semi-vortex components along the $%
x=\pm y$ directions can be anticipated from Eq. (\protect\ref{eps}), showing
a strong change in the spectrum at $k_{x}=\pm k_{y}$.}
\label{f1}
\end{figure}

\begin{figure}[tbp]
\begin{center}
\includegraphics[width=0.45\textwidth]{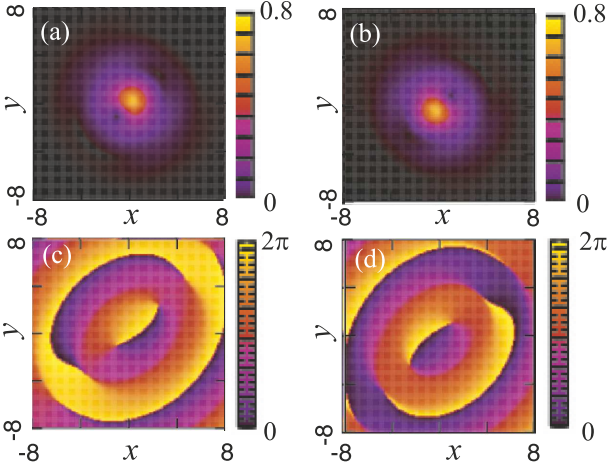}
\end{center}
\caption{Plots of (a) $|\protect\phi _{+}\left( x,y\right)|,$ (b) $|\protect%
\phi _{-}\left( x,y\right)|,$ (c) phase of $\protect\phi_{+}(x,y),$ and (d)
phase of $\protect\phi_{-}(x,y)$ for a mixed-mode state at $\protect\lambda %
_{D}=0.05,$ $\protect\gamma =1.1,$ and $N=3.5$.}
\label{f2}
\end{figure}

In the presence of the Dresselhaus terms, we have found the GS of system (%
\ref{GPERD1}), (\ref{GPERD2}) by means of the imaginary-time simulations. We
start with small $\lambda _{D}=0.05$, and, in particular, focus on the
distinction between the cases of $\gamma <1$ and $\gamma >1$, as they
produce different types of the GS in the absence of the Dresselhaus terms
\cite{Sakaguchi14}. The results demonstrate that stationary solutions keep
the same vorticity structure which is assumed in initial \textit{ans\"{a}tze}
(\ref{SV}) and (\ref{MM}).

Figures \ref{f1}(a) and (b) display plots of $|\phi _{+}\left( x,y\right) |$
and $|\phi _{-}\left( x,y\right) |$ and Figs. \ref{f1}(c) and (d) are plots
of phases of $\phi _{+}(x,y)$ and $\phi _{-}(x,y)$ for an SV state, at $%
\lambda _{D}=0.05$ and $\gamma =0.9$. Similarly, Figure \ref{f2} shows
moduli and phases of $\phi _{+}(x,y)$ and $\phi _{-}(x,y)$ for a MM state at
$\lambda _{D}=0.05$ and $\gamma =1.1$. The well-defined vorticity for $\phi
_{-}$ is shown in Fig.~\ref{f1} (d), while the phase of the mixed-state
structure of $\phi _{+}$ and $\phi _{-},$ as shown in Figs.~\ref{f2}(c) and
(d) , is more complicated. In terms of spectrum (\ref{eps}), the term
accounting for the distortion of the shape of the density distributions is $%
4\lambda _{D}k_{x}k_{y}$. The plots of the SV state are symmetric with
respect to both diagonals $y=\pm x$, while the components of the MM state
are symmetric solely with respect to $y=x$.

Figures \ref{f1}(a),(b) and \ref{f2}(a),(b) demonstrate that an effective
size of the solitons of both the SV and MM types is $\alpha_{1,2}^{-1/2}\sim%
\beta_{1,2}^{-1/2}\sim 4$. Taking into regard the above-mentioned relation
between the scaled and physical units, this implies that the actual size of
the 2D solitons is $\sim 3$ $\mathrm{\mu }$m, i.e., it is comparable to the
typical length of the transverse confinement, which is a generic feature of
matter-wave solitons \cite{Randy}.

\begin{figure}[tbp]
\begin{center}
\includegraphics[width=0.3\textwidth]{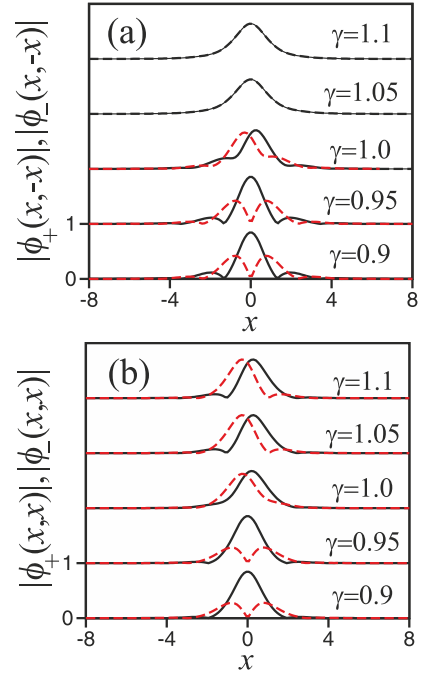}
\end{center}
\caption{(a) Profiles $|\protect\phi _{+}\left( x,y\right) |$ and $|\protect%
\phi _{-}\left( x,y\right) |$ are shown, respectively, by the solid and
dashed lines in diagonal cross section $y=-x$ of the 2D solitons for $%
\protect\gamma =0.90,0.95,1.00,1.05$, and $1.10.$ (b) The same, shown in the
perpendicular diagonal section, $y=x$. Here $\protect\lambda _{D}=0.05$ and $%
N=3.5$. At $\protect\gamma =0.9$ and $\protect\gamma =1.1$ these plots
correspond to Figs. \protect\ref{f1}(a),(b) and \protect\ref{f2}(a),(b),
respectively.}
\label{f3}
\end{figure}

To address in detail the crucially important effect of the switch between
the SV and MM with the increase of the relative cross-attraction strength, $%
\gamma $, Figs. \ref{f3}(a) and (b) display $|\phi _{+}\left( x,y\right) |$
and $|\phi _{-}\left( x,y\right) |$ (solid and dashed lines, respectively),
as produced by the imaginary-time integration, in diagonal cross sections $%
y=\mp x$, for $\gamma =0.90,0.95,1.00,1.05$, and $1.10$. An essential
conclusion is that the switch happens, as in the case of $\lambda _{D}=0$,
exactly at $\gamma =1$, and this critical value does not depend on $\lambda
_{D}$, as long as the 2D solitons exist.

Further, systematic simulations of the evolution of the SV and MM modes with
small arbitrary perturbations added to them (not shown here in detail)
demonstrate that, as well as in the case of $\lambda _{D}=0$, all the
existing SVs and MMs are stable, respectively, at $\gamma <1$ and $\gamma
>1, $ and unstable in the opposite case. The calculation of the system's
energy,
\begin{eqnarray}
E &=&\iint \Bigg\{\frac{1}{2}\left( |\nabla \phi _{+}|^{2}+|\nabla \phi
_{-}|^{2}\right) -  \notag \\
&&\frac{1}{2}\left( |\phi _{+}|^{4}+|\phi _{-}|^{4}\right) -\gamma |\phi
_{+}|^{2}|\phi _{-}|^{2}+  \notag \\
&&\left[ \phi _{+}^{\ast }\left( \widehat{D}^{[-]}-i\lambda _{D}\widehat{D}%
^{[+]}\right) \phi _{-}\right. -  \notag \\
&&\left. \phi _{-}^{\ast }\left( \widehat{D}^{[+]}+i\lambda _{D}\widehat{D}%
^{[-]}\right) \phi _{+}\right] \Bigg\}dxdy,  \label{energy}
\end{eqnarray}%
for both types of the solitons corroborates that, also similar to what is
known in the case of $\lambda _{D}=0$ \cite{Sakaguchi14}, the relation
between the corresponding energies is $E_{\mathrm{SV}}<E_{\mathrm{MM}}$ at $%
\gamma <1$, and vice versa at $\gamma >1$, i.e., the SV and MM, if they
exist, most plausibly realize the system's GS at $\gamma <1$ and $\gamma >1,$
respectively (not shown here in detail).
\begin{figure}[tbp]
\begin{center}
\includegraphics[width=0.3\textwidth]{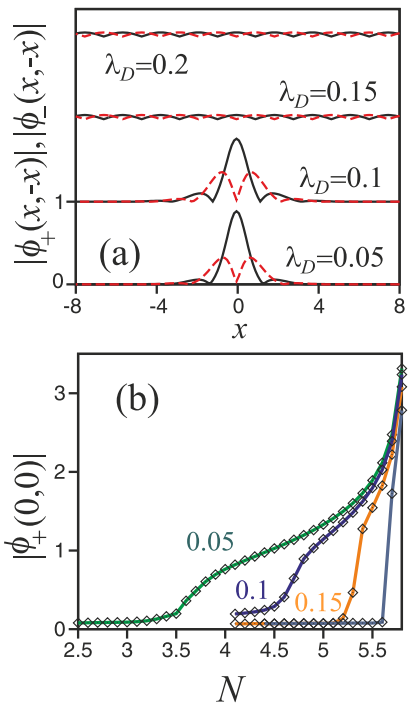}
\end{center}
\caption{(a) Fields $|\protect\phi _{+}\left( x,y\right) |$ and $|\protect%
\phi _{-}\left( x,y\right) |$ in diagonal cross section $y=-x$ at $\protect%
\lambda _{D}=0.05,0.10,0.15$, and $0.20$ for $\protect\gamma =0$ and $N=5$.
(b) The amplitude of component $|\protect\phi _{+}\left( x,y\right) |$
versus $N$ at $\protect\lambda _{D}=0.05,0.10,0.15$ (marked near the plots),
and $0.20$ at $\protect\gamma =0$. }
\label{f4}
\end{figure}

\subsection{Mode-delocalization transition}

The most essential effect caused by the addition of the Dresselhaus SOC
terms is the disappearance of the self-trapped localized modes with the
increase of $\lambda _{D}$ at a critical value, $\lambda _{D}^{\mathrm{(cr)}%
} $, followed by a transition to delocalized states at $\lambda _{D}>$ $%
\lambda _{D}^{\mathrm{(cr)}}$. The growing relative strength of the
Dresselhaus coupling causes the delocalization, rather than the collapse, as
the norm of the solitons remains below the above-mentioned threshold
necessary for the onset of the 2D wave collapse. To illustrate this effect,
Fig. \ref{f4}(a) displays $|\phi _{+}\left( x,y\right) |$ and $|\phi
_{-}\left( x,y\right) |$ in diagonal section $y=-x$ at $\lambda
_{D}=0.05,0.10,0.15$, and $0.20$ for $\gamma =0$ and a fixed norm $N=5$. The

SV solitons exist at $\lambda _{D}=0.05$ and $0.10$, but are replaced by
delocalized states already at relatively small values of the
Dresselhaus-coupling strength, $\lambda _{D}=0.15$ and $0.20$.

A detailed picture of the delocalization transition is provided by Fig. \ref%
{f4}(b), which shows the amplitude (largest value) of $|\phi _{+}\left(
x,y\right) |$ as a function of $N$ at $\lambda _{D}=0.05,0.10,0.15$, and $%
0.20$ for $\gamma =0$. The delocalization is signaled by the drop of the
amplitude to very small values at $N<N_{\min }(\lambda _{D})$ -- for
instance, with $N_{\min }\left( \lambda _{D}=0.05\right) \approx 3.5$. Thus,
the SV solitons exist in the interval of the norm
\begin{equation}
N_{\min }(\lambda _{D})<N<N_{\max }\left( \gamma <1\right) ,  \label{minmax}
\end{equation}%
where the largest norm,
\begin{equation}
N_{\max }\left( \gamma <1\right) \approx 5.85,  \label{Nmax}
\end{equation}%
is the critical value at which the 2D collapse commences in the framework of
the NLSE in free space, hence no solitons can exist at $N>N_{\max }\left(
\gamma <1\right) $. In the limit of $N\rightarrow N_{\max }\left( \gamma
<1\right) $, which corresponds to $\mu \rightarrow -\infty $, the bimodal SV
soliton degenerates into the fundamental Townes soliton in one component,
with $N=N_{\max }\left( \gamma <1\right) $ in it, while the other component
becomes empty. Therefore, $N_{\max }\left( \gamma <1\right) $ does not
depend on $\gamma $ in the interval of $\gamma <1$, where the SV plays the
role of the GS. According to the estimate of actual physical parameters
presented above, $N_{\max }\approx 5.85$ implies that the number of atoms in
the soliton is limited by $\sim 15,000$.

\begin{figure}[tbp]
\begin{center}
\includegraphics[width=0.3\textwidth]{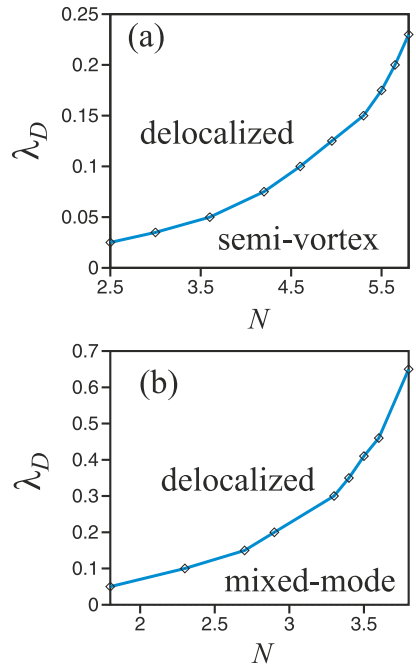}
\end{center}
\caption{Stable semi-vortex and mixed-mode solitons exist in accordingly
labeled regions in the parameter plane of the norm ($N)$ and relative
strength of the Dresselhaus SOC ($\protect\lambda _{D}$), in panels (a) and
(b) for $\protect\gamma =0$ and $\protect\gamma =2$, respectively. The
existence regions for fixed $\protect\lambda _{D}$ correspond to Eq. (%
\protect\ref{minmax}) at $\protect\gamma =0$, and to an accordingly modified
interval at $\protect\gamma =2$, see Eq. (\protect\ref{maxmax}).}
\label{f5}
\end{figure}

The results for both the SV and MM solitons are summarized by Fig. \ref{f5},
which shows the regions in the $\left( N,\lambda _{D}\right) $ planes in
which the solitons exist and are stable ( as has been verified by means of
systematic real-time simulations of their perturbed evolution). Note that,
in panel \ref{f5}(b), the largest norm corresponds to the value (\ref{Nmax})
subjected to obvious rescaling,
\begin{equation}
N_{\max }(\gamma >1)=\frac{2}{1+\gamma }N_{\max }\left( \gamma <1\right) .
\label{maxmax}
\end{equation}%
This is explained by the fact that, in the limit of $N\rightarrow N_{\max
}(\gamma >1)$, which again corresponds to $\mu \rightarrow -\infty $, the
vortical terms vanish in ansatz (\ref{MM}), and the soliton degenerates into
a bound state of two identical Townes solitons in both components. It is
worthy to note that the MM solitons are more immune to the destructive
effect of the Dresselhaus coupling: while the SV modes exist up to $\lambda
_{D}\approx 0.23$ in Fig. \ref{f5}(a), for their MM counterparts the
existence regions extends up to $\lambda _{D}\approx 0.65$, in Fig. \ref{f5}%
(b).

\section{Effect of the Zeeman splitting on 2D solitons}

\subsection{The model}

In this section, we focus on 2D solitons in the SOC system which includes
the Rashba coupling (with scaled strength $\lambda \equiv 1$) and the Zeeman
effect. The latter one breaks the symmetry between the two components of the
pseudo-spinor wave function (the full Rashba-Dresselhaus coupling is also
considered below, in a brief form). The corresponding scaled GPE system is
\begin{gather}
i\frac{\partial \phi _{+}}{\partial t}=-\frac{1}{2}\nabla ^{2}\phi
_{+}-(|\phi _{+}|^{2}+\gamma |\phi _{-}|^{2})\phi _{+}+\widehat{D}^{[-]}\phi
_{-}  \notag \\
-\Omega \phi _{+},  \label{Zeeplus} \\
i\frac{\partial \phi _{-}}{\partial t}=-\frac{1}{2}\nabla ^{2}\phi
_{-}-(|\phi _{-}|^{2}+\gamma |\phi _{+}|^{2})\phi _{-}-\widehat{D}^{[+]}\phi
_{+}  \notag \\
+\Omega \phi _{-},  \label{Zeeminus}
\end{gather}%
where $\Omega $ is the strength of the Zeeman splitting, which is induced,
in the BEC setting, by the optical synthetic magnetic field directed along
the $z-$axis \cite{Lin,Zhai2012,reviews}. The spectrum of the linearized
version of Eqs. (\ref{Zeeplus}), (\ref{Zeeminus}) is%
\begin{equation}
\epsilon _{\pm }=\frac{k^{2}}{2}\pm \sqrt{k^{2}+\Omega ^{2}},  \label{eps2}
\end{equation}%
with a gap $2\Omega $ at $k=0$ (cf. the gapless Rashba-Dresselhaus spectrum
given by Eq. (\ref{eps})). In terms of the estimate for physical parameters
given above, a characteristic strength $\Omega =1$ corresponds, in physical
units, to $\sim $ \ $2\pi \times 100$ Hz for $^{85}$Rb, or \ $2\pi \times 1$
KHz for $^{7}$Li.

Below, we will make use of the system's energy (Hamiltonian), which now has
the form
\begin{eqnarray}
&&E=\iint \Bigg\{\frac{1}{2}\left( |\nabla \phi _{+}|^{2}+|\nabla \phi
_{-}|^{2}\right) -  \notag \\
&&\frac{1}{2}\left( |\phi _{+}|^{4}+|\phi _{-}|^{4}\right) -\gamma |\phi
_{+}|^{2}|\phi _{-}|^{2}-\Omega (|\phi _{+}|^{2}-|\phi _{-}|^{2})  \notag \\
&&+\left[ \phi _{+}^{\ast }\widehat{D}^{[-]}\phi _{-}-\phi _{-}^{\ast }%
\widehat{D}^{[+]}\phi _{+}\right] \Bigg\}dxdy,  \label{E}
\end{eqnarray}%
cf. Eq. (\ref{energy}). Obviously, the increase in $\Omega >0$ should lead
to transfer of atoms from the (pseudo) spin-up component, $\phi _{-}$, to
the spin-down one, $\phi _{+}$, thus attenuating the SOC between the
components and modifying the effect of cross-interaction.

Stationary solutions of Eqs. (\ref{Zeeplus}), (\ref{Zeeminus}) for 2D
solitons with real chemical potential $\mu $ are looked for as $\phi _{\pm
}=\exp \left( -i\mu t\right) u_{\pm }\left( x,y\right) $, where complex
stationary wave functions are determined by equations%
\begin{gather}
\mu u_{+}=-\frac{1}{2}\nabla ^{2}u_{+}-(|u_{+}|^{2}+\gamma |u_{-}|^{2})u_{+}+
\notag \\
\widehat{D}^{[-]}u_{-}-\Omega u_{+},  \label{+} \\
\notag \\
\mu u_{-}=-\frac{1}{2}\nabla ^{2}u_{-}-(|u_{-}|^{2}+\gamma |u_{+}|^{2})u_{-}-
\notag \\
\widehat{D}^{[+]}u_{+}+\Omega u_{-}.  \label{-}
\end{gather}

\subsection{Analytical approaches: large Zeeman splitting and asymptotics of
the SV wave function.}

An analytical approximation can be developed in the limit of large positive $%
\Omega $, when Eq. (\ref{+}) demonstrates that the chemical potential is
close to $-\Omega $:
\begin{equation}
\mu =-\Omega +\delta \mu ,~\left\vert \delta \mu \right\vert \ll \Omega .
\label{mu}
\end{equation}%
The spin-down component, $u_{-}$, being vanishingly small in this limit, Eq.
(\ref{-}) simplifies to%
\begin{equation}
u_{-}\approx \frac{1}{2\Omega }\widehat{D}^{[+]}u_{+},  \label{u-}
\end{equation}%
where $\left( \Omega -\mu \right) $ is replaced by $2\Omega $, pursuant to
Eq. (\ref{mu}). Then, the substitution of approximation (\ref{u-}) into Eq. (%
\ref{+}) leads to the following equation for $u_{+}$:%
\begin{equation}
\left( \delta \mu \right) u_{+}=-\frac{1}{2}\left( 1-\frac{1}{\Omega }%
\right) \nabla ^{2}u_{+}-|u_{+}|^{2}u_{+}~.  \label{deltamu}
\end{equation}%
By itself, Eq. (\ref{deltamu}) is tantamount to the NLSE in the free 2D
space, which gives rise to the Townes solitons; then, Eq. (\ref{u-})
generates a small vortex component of the SV complex. A crucially important
fact is the necessity to scale out factor $\left( 1-1/\Omega \right) $ in
Eq. (\ref{deltamu}). Due to the smallness of $1/\Omega $, the scaling easily
demonstrates that the norm of the SV complex is, in the present case,%
\begin{equation}
N=\left( 1-\frac{1}{\Omega }\right) N_{\max }\left( \gamma <1\right) +%
\mathcal{O}\left( \frac{1}{\Omega ^{2}}\right) ,  \label{<}
\end{equation}%
where the last term is a second-order correction corresponding to the norm
of the small vortex component given by Eq. (\ref{u-}). Thus, Eq. (\ref{<})
(which is compared to the corresponding numerically found dependence below
in Fig. \ref{f6p3}) demonstrates that the total norm of the SV soliton,
produced by the present approximation, is (slightly) \emph{smaller} than the
collapse threshold, $N_{\max }\left( \gamma <1\right).$ For this reason, the
SV soliton remains \emph{protected} against the collapse and stable, still
realizing the GS of the system.

The approximation can be extended to the more general system, when the SOC
includes both the Rashba and Dresselhaus terms, see Eqs. (\ref{GPERD1}),(\ref%
{GPERD2}). In this case, Eq. (\ref{u-}) is replaced by%
\begin{equation}
u_{-}\approx \frac{1}{2\Omega }\widehat{D}^{[+]}u_{+}+\frac{i\lambda _{D}}{%
2\Omega }\widehat{D}^{[-]}u_{+},  \label{Du-}
\end{equation}%
and Eq. (\ref{deltamu}) takes a more general form too:%
\begin{eqnarray}
\left( \delta \mu \right) u_{+}&=&-\frac{1}{2}\left( 1-\frac{1+\lambda
_{D}^{2}}{\Omega }\right) \nabla ^{2}u_{+}+\frac{2\lambda _{D}}{\Omega }%
\frac{\partial ^{2}u_{+}}{\partial x\partial y}  \notag  \label{Ddeltamu} \\
&&-|u_{+}|^{2}u_{+}~.
\end{eqnarray}%
The transformation%
\begin{eqnarray}
x &\equiv &\left( 1-\frac{1+\lambda _{D}^{2}}{2\Omega }\right) \xi -\frac{%
\lambda _{D}}{\Omega }\eta ,  \notag \\
y &\equiv &\left( 1-\frac{1+\lambda _{D}^{2}}{2\Omega }\right) \eta -\frac{%
\lambda _{D}}{\Omega }\xi  \label{xieta}
\end{eqnarray}%
provides for the diagonalization of the linear operator in Eq. (\ref%
{Ddeltamu}), again casting it in the form of the 2D free-space NLSE, and
making it possible to use the Townes soliton as a solution for $u_{+}$, in
terms of coordinates $\left( \xi ,\eta \right) $. Finally, the Jacobian of
transformation (\ref{xieta}) leads to a generalization of Eq. (\ref{<}),%
\begin{equation}
N=\left( 1-\frac{1+\lambda _{D}^{2}}{\Omega }\right) N_{\max }\left( \gamma
<1\right) +\mathcal{O}\left( \frac{1}{\Omega ^{2}}\right) .
\end{equation}%
As well as Eq. (\ref{<}), this result, having $N<N_{\max }\left( \gamma
<1\right) $, secures the protection of the SV soliton against the collapse,
i.e., its stability.

Note that the lowest-order approximation developed here does not give rise
to terms including $\gamma $, and actually implies that, in the limit of
large $\Omega $, all solitons, if they are still stable, belong to the SV
type, irrespective of the value of $\gamma $. This conclusion is consistent
with more general results produced below.

In addition, one can find analytically the asymptotic form of the wave
function in the presence of the Zeeman splitting. For $\Omega =0$, an
asymptotic form of the SV soliton solution to Eqs. (\ref{Zeeplus}), (\ref%
{Zeeminus}) at $r\rightarrow \infty $ was found in Ref. \cite{Sakaguchi14}:%
\begin{eqnarray}
&&\phi _{+}\left( x,y,t\right) =e^{-i\mu t}f_{1}(r),  \notag \\
&&\phi _{-}\left( x,y,t\right) =e^{-i\mu t+i\theta }rf_{2}(r),  \label{frf}
\end{eqnarray}%
with
\begin{eqnarray}
&&f_{1}^{\left( \Omega =0\right) }\approx Fr^{-1/2}e^{-r/R_{\mathrm{SV}%
}}\cos \left( r/L_{\mathrm{so}}+\delta \right) , \\
&&f_{2}^{\left( \Omega =0\right) }\approx -Fr^{-3/2}e^{-r/R_{\mathrm{SV}%
}}\sin \left( r/L_{\mathrm{so}}+\delta \right) ,  \label{asympt}
\end{eqnarray}%
where $F$ and $\delta $ are arbitrary real constants, $R_{\mathrm{SV}}=1/%
\sqrt{-(2\mu +1)}$ is the localization radius of the state, and $L_{\mathrm{%
so}}=1$ (in the present notation) is the spin precession length due to the
spin-orbit coupling. As it follows from condition of real $R_{\mathrm{SV}}$,
the localized modes exist at values of the chemical potential
\begin{equation}
\mu <-1/2.
\end{equation}%
In the presence of a moderately strong Zeeman splitting, with $0<\Omega <1,$
the SV solitons exist at%
\begin{equation}
\mu <-\frac{1+\Omega ^{2}}{2},  \label{mu<}
\end{equation}%
cf. existence region (\ref{mu}) at $\Omega =0$. In this case, the asymptotic
form of the soliton is more complex than one given by Eq. (\ref{asympt}),
with the localization and precession lengths presented as%
\begin{eqnarray}
R_{\mathrm{SV}}^{-1} &=&\sqrt{-(\mu +1)+\sqrt{\mu ^{2}-\Omega ^{2}}},~
\label{-->} \\
L_{\mathrm{so}}^{-1} &=&\frac{\sqrt{-(2\mu +1+\Omega ^{2})}}{\sqrt{-(\mu +1)+%
\sqrt{\mu ^{2}-\Omega ^{2}}}}.
\end{eqnarray}

Strong Zeeman splitting, with $\Omega >1$, replaces existence condition (\ref%
{mu<}) by $\mu <-1$. More specifically, the SV solitons keep the asymptotic
form (\ref{-->}) in the semi-infinite interval (\ref{mu<}) of the chemical
potentials. However, in the additional finite interval appearing in this
case,
\begin{equation}
-\frac{1+\Omega ^{2}}{2}<\mu <-1,  \label{muinterval}
\end{equation}%
the SV soliton exhibits a more dramatic change of its asymptotic shape:
since the Zeeman coupling suppresses the displacement-dependent spin
rotation, the radial oscillations vanishes, while the exponential decay of
the solution at $r\rightarrow \infty $ is provided by%
\begin{equation}
R_{\mathrm{SV}}=\frac{1}{\sqrt{-2\left(\mu+1+\sqrt{2\mu +1+\Omega^{2}}\right)%
}}.  \label{RSV}
\end{equation}
This analytical prediction is compared to numerical results in Fig. \ref%
{f6p1}(b) below.

\begin{figure}[tbp]
\begin{center}
\includegraphics[width=0.3\textwidth]{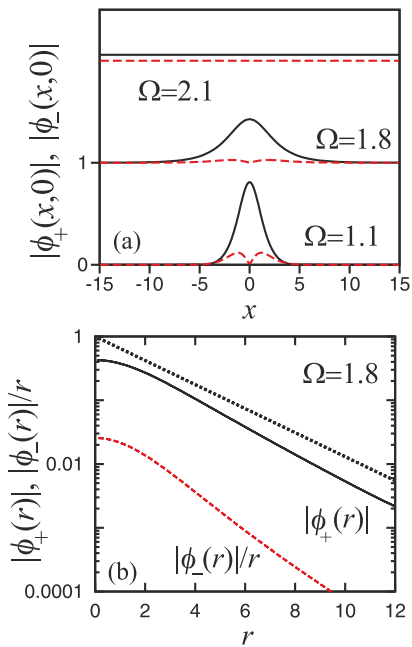}
\end{center}
\caption{(a) Profiles of $|\protect\phi _{+}\left(x,0\right)|$ and $|\protect%
\phi_{-}\left(x,0\right)|$ at $\Omega=1.1,1.8$, and 2.1 for $\protect\gamma %
=0,N=3$. (b) Profiles of $|\protect\phi_{+}(r)|$ and $|\protect\phi%
_{-}(r)|/r $ at $\Omega=1.8$ in the log-scale. The dashed straight line is
the asymptotic exponential form corresponding to Eq. (\protect\ref{RSV}). }
\label{f6p1}
\end{figure}

\subsection{Semi-vortex states}

We begin with the SVs, which, as shown below, are more immune to the action
of the Zeeman splitting than the MM states. First, a VA may be applied,
similar to that developed for the case of $\Omega =0$ in Ref. \cite%
{Sakaguchi14}. Using for this purpose the Gaussian ansatz (\ref{SV}) and
substituting it into expression (\ref{E}) for the system's energy yields the
energy as a function of parameters of the ansatz, $A_{1,2}$ and $%
\alpha_{1,2} $:%
\begin{eqnarray}
&&E_{\mathrm{SV}}=\pi \left[ \frac{A_{1}^{2}}{2}-\frac{A_{1}^{4}}{8\alpha
_{1}}+\frac{A_{2}^{2}}{2\alpha _{2}}-\frac{A_{2}^{4}}{64\alpha _{2}^{3}}-%
\frac{\gamma A_{1}^{2}A_{2}^{2}}{4(\alpha _{1}+\alpha _{2})^{2}}+\right.
\notag \\
&&\left. \Omega \left(\frac{A_{2}^{2}}{4\alpha_{2}^{2}}-\frac{A_{1}^{2}}{%
2\alpha_{1}}\right) +\frac{4A_{1}A_{2}\alpha _{1}}{(\alpha _{1}+\alpha
_{2})^{2}}\right] ,  \label{semivortexVE}
\end{eqnarray}%
with total norm (\ref{N}) expressed as
\begin{equation}
N_{\mathrm{SV}}=\frac{\pi A_{1}^{2}}{2\alpha _{1}}+\frac{\pi A_{2}^{2}}{%
4\alpha _{2}^{2}}.  \label{NSV}
\end{equation}%
Then, the VA predicts values of the four parameters for the SV soliton as a
point at which energy (\ref{semivortexVE}) attains a minimum, subject to
constraint (\ref{NSV}).

%A family of the SV solitons was produced, in parallel, by means of the
%imaginary-time simulations of Eqs. (\ref{Zeeplus}), (\ref{Zeeminus}) and via the VA. The result is
%that, at a fixed value of $N$, the soliton's amplitude decreases, while the
%soliton spreads out, with the increase of $\Omega $. Eventually, the
%amplitude vanishes at some critical point, $\Omega =\Omega _{\mathrm{cr}}$,
%and only delocalized states exist at $\Omega >\Omega _{\mathrm{cr}}$. Figure %
%\ref{f6}(a) displays this trend by showing the amplitude of the larger
%(spin-up) component, $|\phi _{+}(x=0,y=0)|$, as a function of $\Omega $ for $%
%\gamma =0$ and a fixed norm, $N=3$. In this case, the delocalization sets in
%at
%\begin{equation}
%\Omega _{\mathrm{cr}}(N=3)\approx 1.95.  \label{1.95}
%\end{equation}%

\begin{figure}[tbp]
\begin{center}
\includegraphics[width=0.3\textwidth]{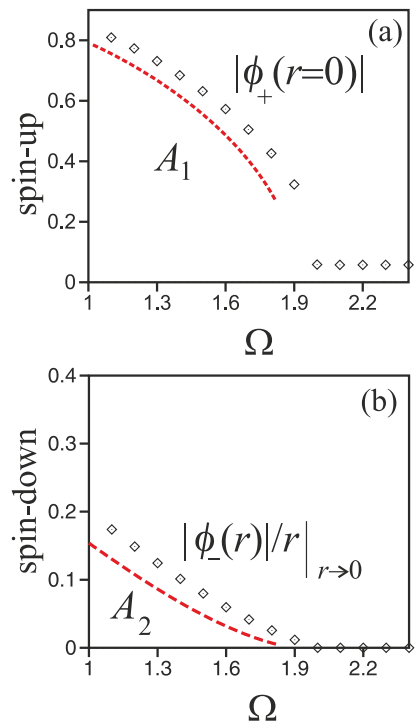}
\end{center}
\caption{(a) The amplitude of the larger (spin-up) component, $|\protect\phi %
_{+}\left(r\right) |$, in the SV state as a function of strength $\Omega $
of the Zeeman splitting, for $\protect\gamma =0$ and $N=3$. (b) The
amplitude of the smaller (spin-down) component, $|\protect\phi%
_{-}\left(r\right) |$, as a function of strength $\Omega $ of the Zeeman
splitting, for $\protect\gamma =0$ and $N=3$. In both panels (a) and (b),
chains of rhombuses and dashed lines show, severally, numerical results and
their counterparts produced by the variational approximation based on the
minimization of energy (\protect\ref{semivortexVE}), subject to constraint (%
\protect\ref{NSV}).}
\label{f6p2}
\end{figure}

%%%%%%%%%%%%%%%%%%%%%%%%%%%%%%%%%%%%%%%%%%%
A family of the SV solitons was produced, in parallel, by means of the
imaginary-time simulations of Eqs. (\ref{Zeeplus}), (\ref{Zeeminus}) and via
the VA. Since the axial symmetry here is preserved, we use the $(r,\theta)$
as well as the $(x,y)$ representation to describe the SV states. Figure \ref%
{f6p1}(a) shows the profiles of $|\phi_{+}\left(r\right)|$ and $%
|\phi_{-}\left(r\right)|$ at $\Omega=1.1,1.8$, and 2.1 for $\gamma =0,N=3$.
For $\Omega=1.8$, $\mu$ satisfies the condition $-(1/2)(1+\Omega^2)<\mu<-1$.
Figure \ref{f6p1}(b) compares the profile of $|\phi_{+}\left(r\right)|$ with
the asymptotic form in Eq. (\ref{RSV}). The result is that, at a fixed value
of $N$, the soliton's amplitude decreases, while the soliton spreads out,
with the increase in $\Omega$. Eventually, the amplitude vanishes at some
critical field, $\Omega =\Omega _{\mathrm{cr}}$, and only delocalized states
exist after this threshold. Figures \ref{f6p2}(a) and (b) display this trend
by showing the amplitudes of (a) the larger (spin-up) component, $%
|\phi_{+}(r)|,$ and (b) the smaller (spin-down) component, $|\phi _{-}(r)|$,
as a function of $\Omega $ for $\gamma =0$ and a fixed norm $N=3$. In this
case, the delocalization sets in at
\begin{equation}
\Omega _{\mathrm{cr}}(N=3)\approx 1.95.  \label{1.95}
\end{equation}%
It is also seen that the VA provides a reasonable accuracy, predicting, in
particular, $\Omega _{\mathrm{cr}}^{\mathrm{(VA)}}(N=3)\approx 1.83$, with
relative error $\approx 6\%$. %%%%%%%%%%%%%%%%%%%%%%%%%%%%%%%%%%%%%%%%%%%

\begin{figure}[tbp]
\begin{center}
\includegraphics[width=0.3\textwidth]{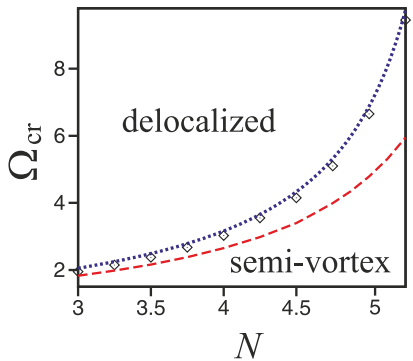}
\end{center}
\caption{The critical value $\Omega_{\mathrm{cr}}$, up to which the
semi-vortex soliton persists, versus its norm $N,$ for $\protect\gamma =0$.
Chains of rhombuses and dashed lines show, correspondingly, numerical
results and their counterparts produced by the variational approximation
based on the minimization of energy. The dotted line is the prediction of
Eq.~(\protect\ref{<}).}
\label{f6p3}
\end{figure}

Figure \ref{f6p3} produces a summary of the numerical and variational
results for the SV solitons, showing the existence region for stable SVs in
the parameter plane of $\left( N,\Omega \right) $ for $\gamma =0$. Although
the plot is confined to $N\leq 5.25$, the analytical result given above by
Eqs. (\ref{u-})-(\ref{<}) suggests that the SV existence boundary in Fig. %
\ref{f6p3}(b) extends to $\Omega \rightarrow \infty $ in the limit of $%
N\rightarrow N_{\max }\left( \gamma <1\right) \approx 5.85$.

\subsection{Mixed-mode states and mixed-mode - semi-vortex transitions}

\begin{figure}[tbp]
\begin{center}
\includegraphics[width=0.3\textwidth]{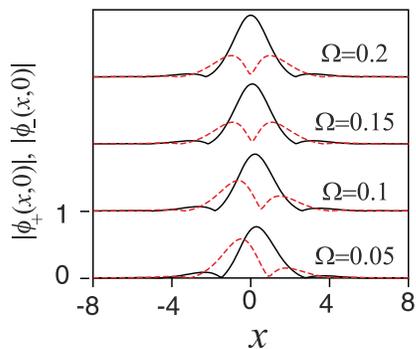}
\end{center}
\caption{Profiles of the spin-up and spin-down components, $|\protect\phi %
_{+}\left( x,0\right) |$ and $|\protect\phi _{-}\left( x,0\right) |$ (shown
by continuous and dashed lines, respectively) of mixed-mode solitons at the
Zeeman splitting $\Omega=0.05,0.10,0.15,$ and $0.20$, for $\protect\gamma %
=1.5$ and $N=3$. Eventually, the mixed mode transforms into a semi-vortex.}
\label{f7}
\end{figure}

As well as in the case of the SVs, the increase of strength $\Omega $ of the
Zeeman splitting leads to the reduction of the amplitude of the spin-down
component $\phi _{-}$ of the MM soliton, in comparison with its spin-up
counterpart, $\phi _{+}$, as shown in Fig. \ref{f7}. However, it is also
observed in Fig. \ref{f7} that, instead of the delocalization, the MM
undergoes a transformation into a stable SV soliton -- even at $\gamma >1$,
when solely the MM states, but not SVs, may be stable in the absence of the
Zeeman splitting.
\begin{figure}[tbp]
\begin{center}
\includegraphics[width=0.3\textwidth]{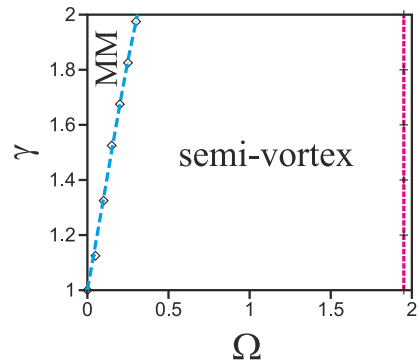}
\end{center}
\caption{The chain of rhombuses shows the numerically found cross-attraction
strength, $\protect\gamma $, at which the ground state switches form the
mixed mode to the semi-vortex, as a function of the Zeeman $\Omega$ for the
norm $N=3$. The dashed line is the same dependence as predicted by the
variational approximation, see Eq. (\protect\ref{EE}). The ground state is
the semi-vortex beneath this line, and the mixed mode above it. The vertical
line at $\Omega \approx 1.95$ is the Zeeman coupling at which the
semi-vortex with $N=3$ suffers the delocalization, see Eq. (\protect\ref%
{1.95}).}
\label{f8}
\end{figure}

Thus, it is relevant to identify the shift of the MM-SV conversion from
point $\gamma =1$, which was the universal boundary between the SV and MM
shapes of the GS at $\Omega =0$, to $\gamma >1$. This important
characteristic of the system can be predicted by means of the VA, using
ansatz (\ref{MM}) as an approximation for the MM state. The substitution of
the ansatz into Eqs. (\ref{E}) and (\ref{N}) yields the following
expressions, cf. Eqs. (\ref{semivortexVE}) and (\ref{NSV}):
\begin{eqnarray}
&&E_{\mathrm{MM}}=\pi \left[ B_{1}^{2}+\frac{B_{2}^{2}}{\beta _{2}}%
-(1+\gamma )\left( \frac{B_{1}^{4}}{4\beta _{1}}+\frac{B_{2}^{4}}{32\beta
_{2}^{3}}\right) \right.  \notag \\
&&-\left. \frac{B_{1}^{2}B_{2}^{2}}{(\beta _{1}+\beta _{2})^{2}}+\frac{%
8B_{1}B_{2}\beta _{1}}{(\beta _{1}+\beta _{2})^{2}}\right] ,  \label{mixedVE}
\end{eqnarray}%
\begin{equation}
N=\frac{\pi B_{1}^{2}}{\beta _{1}}+\frac{\pi B_{2}^{2}}{2\beta _{2}^{2}}.
\label{NMM}
\end{equation}%
Then, parameters of the MM solitons are predicted by the present version of
the VA as those at which energy (\ref{mixedVE}) attains a minimum, subject
to the constraint of keeping the fixed norm, as per Eq. (\ref{NMM}). Note,
in particular, that this approximation for the energy, produced by ansatz (%
\ref{MM}), does not contain $\Omega $ (in contrast with its SV counterpart (%
\ref{semivortexVE})) because ansatz (\ref{MM}) implies equal norms of $\phi
_{+}$ and $\phi _{-}$, hence the respective expectation value, which
determines the Zeeman energy, vanishes:

\begin{equation}
\left\langle \sigma _{z}\right\rangle =\frac{1}{N}\iint \left( \left\vert
\phi _{+}\right\vert ^{2}-\left\vert \phi _{-}\right\vert ^{2}\right) dxdy=0.
\end{equation}%
This circumstance suggests that the SV state may provide lower energy in the
presence of the Zeeman splitting, thus realizing the GS even at $\gamma >1$.

The VA predicts the transition from the MM to SV as a point at which the
minima of energies predicted by Eqs. (\ref{semivortexVE}) and (\ref{mixedVE}%
) become equal, for a given norm,
\begin{equation}
\min \left\{ E_{\mathrm{MM}}(N)\right\} =\min \left\{ E_{\mathrm{SV}%
}(N)\right\} .  \label{EE}
\end{equation}%
The result, in the form of the $\gamma (\Omega )$ dependence following from
Eq. (\ref{EE}), along with its numerically generated counterpart, is
displayed in Fig. \ref{f8}, which demonstrates that the VA provides for a
very accurate prediction of the MM $\rightarrow $ SV transition point.

The vertical dashed line in Fig. \ref{f8}, which bounds the existence area
of the SVs, corresponds to the critical value of $\Omega $, given by Eq. (%
\ref{1.95}), at which the SV with norm $N=3$ suffers delocalization.
Although this $\Omega _{\mathrm{cr}}$ was obtained above for $\gamma =0$, it
actually pertains to all values of $\gamma $, because, as seen from Fig. \ref%
{f6p2}, close to the delocalization transition, the amplitude of the $%
\phi_{-}$ component becomes negligibly small in comparison with that of $%
\phi _{+}$, hence the cross-interaction term is also negligible at the
delocalization point, in comparison with its self-interaction counterpart.

\section{Conclusion}

In this paper, we have extended the analysis for 2D solitons in the
pseudo-spinor BEC with attractive nonlinearity, which may be stabilized in
the form of SV (semi-vortex) and MM (mixed-mode) localized states by the SOC
(spin-orbit coupling). We have considered the generic case of the combined
Rashba-Dresselhaus SOC and analyzed the effect of the Zeeman splitting in
the presence of the Rashba coupling. Families of SV and MM solitons have
been constructed by means of numerical and approximate analytical methods,
the largest number of atoms possible in the solitons under physically
relevant conditions being $\sim 1.5\times10^{4}$, while as characteristic
size of the solitons is $\sim 3$ $\mathrm{\mu }$m. The increase in the
strength of the Dresselhaus coupling preserves the soliton type (SV or MM)
and eventually leads its delocalization. The sufficiently strong Zeeman
splitting converts the MM solitons into the SV ones, which also eventually
suffer delocalization. The existence regions have been found for both
soliton species. These results help to understand novel possibilities for
the creation of stable vorticity-bearing solitons in matter-wave settings,
offered by the introduction of the SOC in its generic form and the synthetic
Zeeman splitting.

As an extension of the present analysis, it may be interesting to consider
mobility of the stable solitons in the present system with broken Galilean
invariance. A challenging possibility is to study effects of the Zeeman
splitting on metastable 3D SOC-supported solitons which were recently found
in Ref. \cite{HP}. More generally speaking, the present results may find
their application for the stabilization of 2D solitons in other nonlinear
models.

\section*{Acknowledgments}

B.A.M. appreciates hospitality of the Interdisciplinary Graduate School of
Engineering Sciences at the Kyushu University (Fukuoka, Japan), and of
Department of Physical Chemistry of the University of the Basque Country.
E.Y.S. acknowledges support of the University of the Basque Country UPV/EHU
under program UFI 11/55, Spanish MEC (FIS2012-36673-C03-01 and
FIS2015-67161-P), and Grupos Consolidados UPV/EHU del Gobierno Vasco
(IT-472-10).

%Unused bibitems

%\bibitem{Abdullaev03} {F. Kh. Abdullaev, J. G. Caputo, R. A. Kraenkel, and
%B. A. Malomed}, {Phys. Rev. A} \textbf{67}, {013605} {(2003)}.
%
%\bibitem{Zhang} Yongping Zhang, Li Mao, and Chuanwei Zhang Phys. Rev. Lett.
%108, 035302 (2012)
%

\end{document}